\documentclass[aps,prb,floatfix,preprint]{revtex4-1}

\usepackage{amsmath,amssymb}
\usepackage{natbib}
\usepackage{graphicx}
\usepackage{dcolumn}
\usepackage{bm}
\usepackage{subfigure}

\usepackage[pdffitwindow=false,bookmarks=true,pdfauthor={Hoffmann et al.},colorlinks=true,citecolor=blue,linkcolor=blue]{hyperref}
\usepackage{hypcap}

\begin{document}


\title{CO oxidation on Pd(100) vs. PdO\texorpdfstring{($\sqrt{5}\times \sqrt{5})R27^{\circ}$}{sqrt5xsqrt5 R27}:\\
First-Principles Kinetic Phase Diagrams and Bistability Conditions}

\author{Max J. Hoffmann}
\affiliation{Department Chemie, Technische Universit{\"a}t M{\"u}nchen,
  Lichtenbergstr. 4, D-85747 Garching, Germany}

\author{Karsten Reuter}
\affiliation{Department Chemie, Technische Universit{\"a}t M{\"u}nchen,
  Lichtenbergstr. 4, D-85747 Garching, Germany}

\begin{abstract}
We present first-principles kinetic Monte Carlo (1p-kMC) simulations addressing the CO oxidation reaction at Pd(100) for gas-phase conditions ranging from ultra-high vacuum (UHV) to ambient pressures and elevated temperatures. For the latter technologically relevant regime there is a long-standing debate regarding the nature of the active surface. The pristine metallic surface, an ultra-thin $(\sqrt{5}\times \sqrt{5})R27^{\circ}$PdO(101) surface oxide, and thicker oxide layers have each been suggested as \emph{the} active state. We investigate these hypotheses with 1p-kMC simulations focusing on either the Pd(100) surface or the PdO(101) surface oxide and intriguingly obtain a range of $(T,p)$-conditions where both terminations appear metastable. The predicted bistability regime nicely ties in with oscillatory behavior reported experimentally by Hendriksen and coworkers [Catal. Today {\bf 105}, 234 (2005)]. Within this regime we find that both surface terminations exhibit very similar intrinsic reactivity, which puts doubts on attempts to assign the catalytic function to just one \emph{active} state.
\end{abstract}

\maketitle

\section{Introduction}

Common use of Pd for catalytic exhaust gas pu\-ri\-fi\-ca\-tion \cite{gandhi_automotive_2003} has motivated frequent studies on Pd single-crystal model catalysts to elucidate the underlying molecular-level mechanisms. Notwithstanding, despite significant efforts the surface structure and composition at \emph{near-ambient} conditions remains unclear. This is largely due to inherent difficulties in achieving atomic-scale information/resolution in this technologically relevant regime of near-ambient pressures and at temperatures of \mbox{300-800}~K. Generally, the outcome of such experiments seems strongly dependent on the preparation conditions and experimental setup. Specifically, there exists a longstanding and controversial debate on whether reactivity is due to the pristine metal surface\cite{gao_reply_2010}, a surface oxide film\cite{van_rijn_comment_2010}, or even a thicker bulk-like oxide overlayer.\cite{hirvi_co_2010}

Under ultra-high vacuum (UHV) conditions, different oxidation stages of Pd(100) \cite{zheng_reactivity_2002} have been thoroughly characterized both experimentally and theoretically. By low-energy electron diffraction (LEED) Chang and Thiel first identified five distinct ordered structures before the onset of bulk oxide formation: a $p(2\times2)$ oxygen adlayer, a $c(2\times2)$ adlayer, a $(5\times5)$, and a $(\sqrt{5}\times\sqrt{5})R27^{\circ}$ (for brevity henceforth denoted as $\sqrt{5}$) reconstruction.\cite{chang_oxygen_1988} Todorova \emph{et al.}\cite{todorova_pd1_2003} established that the latter $\sqrt{5}$ structure corresponds to a single-layer of PdO(101) on top of Pd(100).

At low pressures several experimental studies concluded that formation of this surface oxide is accompanied with a low catalytic activity.\cite{kiss_deactivation_1985,gabasch_comparison_2007} This was primarily attributed to the low CO binding energy at the surface oxide \cite{zheng_reactivity_2002}, which Gao \emph{et al.} estimated to be around \mbox{0.5-0.6}~eV.\cite{gao_infrared_2008} At elevated pressures the situation is less clear. Lundgren \emph{et al.}\cite{lundgren_kinetic_2004} compared structural information from {\em in situ} surface x-ray diffraction measurements in a pure oxygen environment to a systematic {\em ab initio} surface phase diagram \cite{rogal_first-principles_2007,rogal_co_2008}, and concluded that the formation of bulk oxide on Pd(100) is kinetically severely hindered even at temperatures around 675~K. This suggests that at the $(T,p)$-conditions relevant for near-ambient CO oxidation the most likely surface terminations are either pristine Pd(100) with some coverage of O adsorbates, or a $\sqrt{5}$ monolayer. On top of this, reactor scanning tunneling microscopy (STM) measurements by Hendriksen \emph{et al.} in fact suggested bi\-stability in this near-ambient regime. \cite{hendriksen_oscillatory_2004,hendriksen_bistability_2005,hendriksen_role_2010} In the understanding of the work by Lundgren {\em et al.} this result refers to $(T,p)$ conditions where both O@Pd(100) and the $\sqrt{5}$ surface oxide are metastable.

Here, we investigate this hypothesis with first-principles kinetic Monte Carlo (1p-kMC) simulations that focus on either of the two surface states, i.e. either O@Pd(100) or the $\sqrt{5}$ surface oxide, thereby extending previous 1p-kMC work that had focused exclusively on the $\sqrt{5}$ phase \cite{rogal_co_2007,rogal_first-principles_2007}. We indeed find a range of $(T,p)$-conditions where both surface models appear metastable. Moreover, in this bistability regime, both surface terminations yield roughly similar turnover frequencies (TOFs) for near-ambient pressures. In contrast and consistent with experiment we find the Pd(100) model to be the much more reactive one in the UHV regime. This puts experiments \cite{gao_reply_2010,van_rijn_comment_2010} into perspective that claim one surface state to be \emph{the} active one, irrespective of the specific gas-phase conditions.

\section{Methods}

\subsection{1p-kMC simulations}

Our central goal is to describe the range of gas-phase conditions in which the surface termination of the Pd(100) model catalyst is characterized by pristine Pd(100) or by the $\sqrt{5}$ surface oxide. In order to also account for stabilization effects through the ongoing reaction kinetics this requires analyzing the steady-state coverages of CO and O in microkinetic models of these two surface terminations as a function of external feed conditions $(T, p_{\mathrm{CO}}, p_{\mathrm{O}_2})$. For this, two factors indicate that standard microkinetic modeling in terms of mean-field rate equations will not be sufficient: On pristine Pd(100) lateral interactions between adsorbed O and CO are known to be rather strong\cite{zhang_accuracy_2007,behm_adsorption_1980,liu_chemical_2006,liu_atomistic_2009}, while on the $\sqrt{5}$ surface oxide it is its essentially one-dimensional geometric trench structure that will lead to adlayer inhomogeneities. We thus opt for 1p-kMC simulations as presently only technique that provides the desired microkinetic information while fully accounting for the correlations, fluctuations, and explicit spatial distributions of the chemicals at the catalyst surface.\cite{temel_2007,reuter_2011_book,sabbe_2012}

In 1p-kMC the time evolution of the system is coarse-grained to the discrete rare-event dynamics. Relying on a Markov approximation rejection-free 1p-kMC algorithms thus generate state-to-state trajectories that in their average yield the probability-density function $P_i(t)$ to find the system at time $t$ in state $i$ representing the corresponding potential energy surface (PES) basin $i$. The propagation of this probability density is governed by the Markovian master equation,
   \begin{equation}
   \frac{dP_i(t)}{dt} \;=\; - \sum_{j\ne i} k_{ij} P_i(t) \;+\; \sum_{j\ne i} k_{ji} P_j(t) \quad ,
   \label{master}
   \end{equation}
where the sums run over all system states $j$, and $k_{ij}$ is the rate constant for going from state $i$ to $j$. The central first-principles ingredients required for the 1p-kMC simulation are thus the individual rate constants of the considered elementary processes (adsorption, desorption, diffusion, reaction). To keep this input tractable, i.e. to arrive at a finite number of inequivalent processes and corresponding rate constants, 1p-kMC simulations are commonly performed on lattice models. In the following sections we will first provide the working equations to determine the first-principles rate constants $k_{ij}$ and then detail the specific lattice models employed in the present work for CO oxidation at pristine Pd(100) and at the $\sqrt{5}$ surface oxide.

For given environmental (gas-phase) conditions (which specify the adsorption rate constants), the output of 1p-kMC simulations are then the detailed surface composition and occurrence of each individual elementary process at any time. Since the latter comprises the surface reaction events, this also gives the catalytic activity in form of products per surface area and time (i.e. TOFs), either time-resolved, e.g. during induction, or time-averaged during steady-state operation.

\subsection{First-principles rate constants}

The first-principles rate constants in this work are evaluated following the approach put forward by Reuter and Scheffler \cite{reuter_first-principles_2006}. In brief, this approach relies on (harmonic) transition state theory (TST) for bound to bound processes like diffusion, and kinetic gas theory together with detailed balance to calculate adsorption and desorption rate constants. As the approach and its derivation have been detailed before, we here restrict ourselves to the presentation of the working equations for self-containment.

The adsorption rate constant for species $i$ is given by the rate with which these particles impinge on the unit-cell surface area $A_{uc}$ and the local sticking coefficient $\widetilde{S}_{st,i}(T)$, which gives the fraction of the impinging particles that actually stick to a given free site $st$ at temperature $T$
   \begin{eqnarray}
   k^{\mathrm{ad}}_{st,i}(T, p_i)
   &=& \widetilde{S}_{st,i}(T)\frac{p_i A_{uc}}
   {\sqrt{2\pi m_{i} k_{B}T}} \quad .
   \end{eqnarray}
Here, $k_{B}$ is the Boltzmann constant, $p_i$ the partial pressure of species $i$, and $m_i$ the particle mass. In unactivated adsorption events the local sticking coefficient merely accounts for the number of inequivalent sites in the surface unit-cell ({\em vide infra}). For activated adsorption events or Eley-Rideal (ER) type CO oxidation events, it is additionally governed by the adsorption or reaction barrier $\Delta E_{i,st,j}$, respectively,
   \begin{equation}
   \widetilde{S}_{st,i}(T) \;=\; \left(\frac{A_{st, i}}{A_{uc}}\right)\exp \left(-\frac{\Delta E_{i,st,j}}{k_{B}T}\right) \quad .
   \end{equation}
where $A_{st, i}$ is a geometrical factor reflecting the relative share with which molecules impinge on the different inequivalent surface sites $st$.

Desorption of a particle adsorbed on a surface site $st$ is modeled as the time reversed process of adsorption, and its rate constant thus has to fulfill detailed balance or microscopic reversibility.
   \begin{eqnarray}
   \frac{k^{\mathrm{ad}}_{st,i}(T, p_i)}{k^{\mathrm{des}}_{st,i}(T)}
   &=&\exp \left(\frac{\Delta G_{st,i}(T,p_i)}{k_{B}T} \right)\nonumber\\
   &\approx&
   \exp \left(\frac{ \mu_{\mathrm{gas},i}(T, p_i)-E^{\mathrm{bind}}_{st, i}}{k_{B}T}\right)
   \end{eqnarray}
where $\Delta G_{st,i}(T,p_{i})$ is the difference in Gibbs free energy between the particle adsorbed at the surface state and in the gas phase. This is approximated by the difference between the gas-phase chemical potential $\mu_{\mathrm{gas},i}(T,p_{i})$ \cite{reuter_composition_2001} and the binding energy of the particle in the adsorbed state $E^{\mathrm{bind}}_{st, i}$. Following the procedure detailed in Ref. \onlinecite{reuter_composition_2001} we interpolate tabulated values \cite{chase_jr_nist-janaf_1998} to determine the gas-phase chemical potentials at any gas-phase condition.

The diffusion rate constant of an adsorbate from one surface site $st$ to another site $st'$ is approximated as
   \begin{equation}
   k^{\mathrm{diff}}_{st,st'\!\!,i}(T)
   \approx \left(\frac{k_B T}{h}\right)
   \exp\left(-\frac{\Delta E^{\mathrm{diff}}_{st,st'\!\!,i}}{k_{B} T} \right) \quad ,
   \end{equation}
where $\Delta E^{\mathrm{diff}}_{st,st'\!\!,i}$ is the diffusion energy barrier. Langmuir-Hinshelwood (LH) type CO oxidation reactions are described with an equivalent expression containing the reaction barrier. They, as well as the ER reactions, are generally treated as associative desorption events though, i.e. the formed CO$_2$ immediately desorbs from the surface.

Within this approach the required first-principles input to determine the rate constants reduces to the binding energies of the species at the surface sites (for the desorption rate constant), as well as to their diffusion and reaction barriers. The latter require the determination of the corresponding transition states, which is at present commonly achieved with standard transition state search algorithms like nudged-elastic band \cite{henkelman,henkelman2}. Unfortunately, we found the low PES corrugation for lateral shifts of the $\sqrt{5}$ overlayer \cite{todorova_pd1_2003,kostelnik} to severely affect the performance and reliability of such methods. We therefore instead identified the transition states through scans along suitable reaction coordinates. For diffusion barriers the specific reaction coordinate employed was the lateral coordinate along a straight line connecting the known initial and final state. For CO oxidation reactions the distance between the C atom of the CO molecule and the O adsorbate was employed. We carefully checked for hysteresis effects and estimate the uncertainty in the determined barriers to be of the order of $\pm 0.1$\,eV. The effect of this energetic uncertainty on the presented 1p-kMC results will be critically discussed below and is not found to affect any of our conclusions.

All required total energies were obtained from density-functional theory (DFT) with the generalized gradient approximation functional by Perdew, Becke, and Ernzerhof (PBE) \cite{perdew_generalized_1996} to treat electronic exchange and correlation. Using the plane-wave code CASTEP \cite{clark_first_2005} with standard library ultrasoft pseudopotentials systematic convergence tests showed that the quantities of interest (binding energies, diffusion and reaction barriers) are converged to within 30 meV at the employed energy cut-off of 400~eV and $k$-point density of 0.4~\texorpdfstring{\AA$^{-1}$}{angstrom}. The surfaces were modeled in periodic supercell geometries, containing vacuum separations of more than 10~{\AA}. For the $\sqrt{5}$ surface oxide the calculations were done within a $(1\times 1)$ surface unit-cell of surface oxide on top of four layers of Pd(100). For pristine Pd(100) we employed a $(2 \times 2)$ surface unit-cell and also four metal layer slabs. In both cases, all geometries were fully relaxed to residual forces below 50~meV/{\AA}, while keeping the bottom two slab layers fixed at their bulk positions.

\subsection{1p-KMC lattice models}

\begin{figure}
\includegraphics[width=\columnwidth]{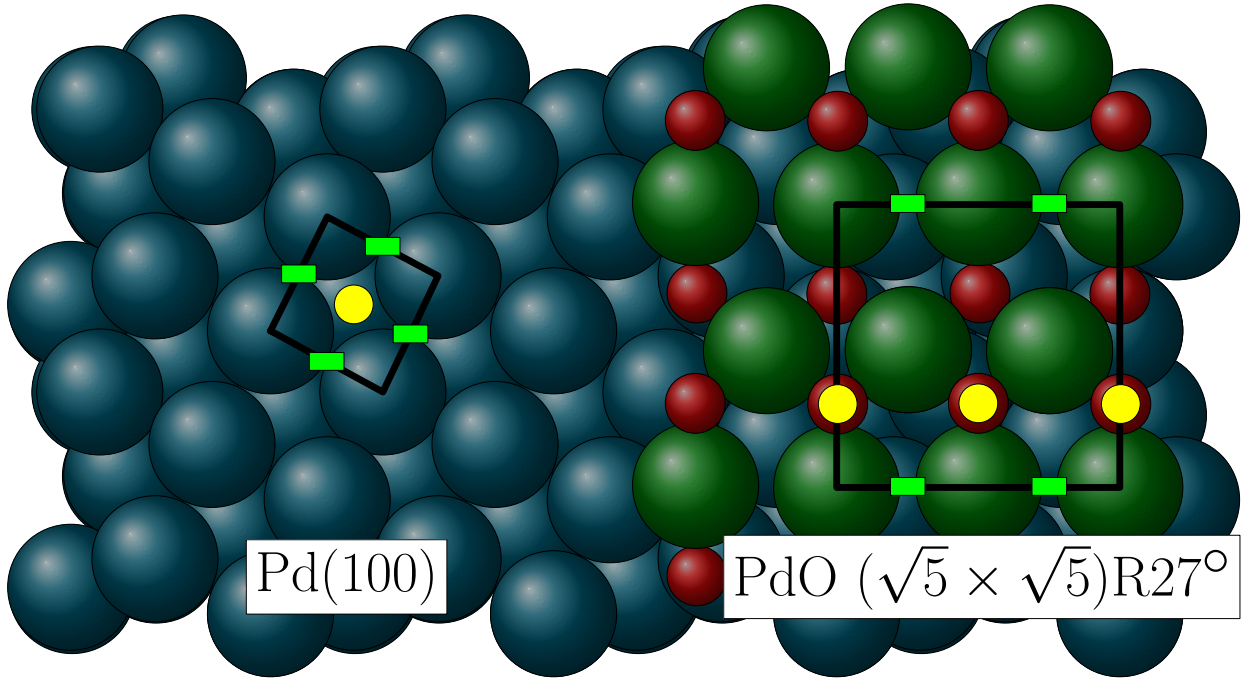}
\caption{\label{geom}
(Color online) Top view of Pd(100) and the $\sqrt{5}$ surface oxide, illustrating the employed lattice models. Hollow sites are depicted by yellow circles and bridge sites by green rectangles (see text). Pd(100) top-layer atoms are represented by large dark green spheres, Pd atoms in the $\sqrt{5}$ oxide by large light green spheres, and O atoms by small red spheres.}
\end{figure}

\subsubsection{Pd(100)}

In setting up the lattice model for Pd(100) we exploit the limited coverage range, for which the 1p-kMC simulations need to provide a faithful representation. Detailed experimental work \cite{chang_oxygen_1988} indicates the formation of surface oxides at around a critical O coverage of 0.5 monolayer (ML, defined as the ratio of O to top layer Pd atoms). As the purpose of this work is to derive the gas-phase conditions under which {\em no} surface oxide is formed, the developed model only targets the coverage range $\theta < 0.5$\,ML and the elementary processes occurring in it. Similarly, CO coverages of more than 0.5\,ML have only been characterized for strongly overstoichiometric CO pressures and low temperatures\cite{behm_adsorption_1980,stuve_co_1984,uvdal_structure_1988,tushaus_understanding_1990}. As this regime is outside the catalytic context, only CO coverages below 0.5\,ML are of interest here, too.

In this coverage range O atoms bind preferably at the fourfold hollow sites at Pd(100) \cite{rieder_helium_1985,zhang_accuracy_2007}, while CO binds at twofold bridge sites \cite{rogal_co_2008,bradshaw_chemisorption_1978}. In consequence we set up the lattice model shown in Fig. \ref{geom}, in which O can exclusively occupy hollow sites and CO bridge sites. Lateral interactions between oxygen adatoms have been systematically calculated by Zhang, Blum, and Reuter and were found to be strongly repulsive at nearest and moderately repulsive at next-nearest hollow-hollow distance \cite{zhang_accuracy_2007}. For the targeted coverage regime and intermediate temperatures up to 600\,K, this is taken into account in the 1p-kMC model by blocking adsorption or diffusion events into configurations that would result in O atoms in nearest-neighbor positions. For dissociative adsorption this implies a required motif of eight empty sites (two next-nearest neighbor sites for the actual adsorption and their immediately adjacent six nearest-neigbor sites) known as the 8-site rule\cite{brundle_summary_1984,chang_formation_1987,liu_atomistic_2006,liu_atomistic_2009} that is required for an adsorption event to take place. 

Similarly repulsive interactions between adsorbed CO molecules are indicated by the experimentally characterized superstructures \cite{behm_adsorption_1980,stuve_co_1984,uvdal_structure_1988,tushaus_understanding_1990} and DFT calculations \cite{rogal_first-principles_2007}. Accordingly and accounting for the shorter bridge-bridge distances, we also block CO adsorption and diffusion events into configurations resulting in CO molecules up to next nearest-neighbor bridge-bridge distance. We note, however, that the next nearest-neighbor lateral repulsion is not too strong \cite{liu_atomistic_2006}, and might hence be overestimated by the simple site-blocking in our default model, henceforth denoted as Pd(100)-2NN. We assess this below by additionally considering a model, in which CO-CO coadsorption is blocked only at nearest bridge-bridge distance (Pd(100)-1NN) and find this not to affect our conclusions towards bistability.

Finally, the absence of mixed O-CO adsorbate structures at low temperatures \cite{stuve_co_1984} suggests O-CO lateral interactions to also be repulsive. Unfortunately, no systematic first-principles studies characterizing these interactions have been performed to date. In this situation we note that the nearest-neighbor hollow-bridge distance at Pd(100) is smaller than the O-CO distance in the CO oxidation transition states described below. O-CO lateral interactions are thus modeled by blocking adsorption and diffusion events into configurations involving O and CO at such close distance.

Within this lattice model and site-blocking rules we consider all non-concerted O and CO adsorption, desorption, as well as nearest-neighbor site diffusion and reaction events. Oxygen adsorption is modeled as dissociative process into two vacant next-nearest hollow sites. CO adsorption occurs unimolecularly into one free bridge site. Both adsorption processes are treated as non-activated, i.e. using a local sticking coefficient of 1 for the dissociative adsorption of O$_{2}$ and 0.5 for CO, the latter accounting for the two bridge sites per surface unit-cell. For oxygen adsorption in the low-coverage range this is supported by explicit DFT sticking coefficient calculations \cite{meyer_first-principles_2011}, while for CO we verified that lifting the CO molecule vertically up from the bridge adsorption site yields a path without adsorption barrier. This together with the observation that a CO lowered above an adsorbed oxygen atom is efficiently steered into a neighboring vacant bridge site, supports the assumption that all CO molecules impinging in the vicinity of a vacant bridge site will stick.

    \begin{table}
    \label{dft_energetics}
    \caption{Summary of DFT binding energies, diffusion and reaction barriers (LH and ER) used in the Pd(100) and $\sqrt{5}$ 1p-kMC
    models. All values are in eV.}
    \begin{ruledtabular}
      \begin{tabular}{llll}
      \multicolumn{4}{l}{PdO$\sqrt{5}$ desorption barriers:}\\
      \multicolumn{4}{l}{$E^0$ on-site energy, $V$ nearest-neighbor lateral interaction$\footnotemark[1]$}\\
      \hline
      $\stackrel{E^0_{\tiny{\mathrm{O, bridge}}}}{-0.51\footnotemark[1]}$ &
      $\stackrel{E^0_{\tiny{\mathrm{O, hollow}}}}{-1.95\footnotemark[1]}$ &
      $\stackrel{E^0_{\tiny{\mathrm{CO, bridge}}}}{-1.40\footnotemark[1]}$ &
      $\stackrel{E^0_{\tiny{\mathrm{CO, hollow}}}}{-1.92\footnotemark[1]}$ \\

      $\stackrel{V_{\tiny{\mathrm{O-O,br-br}}}}{0.08\footnotemark[1]}$ &
      $\stackrel{V_{\tiny{\mathrm{O-O,hol-hol}}}}{0.07\footnotemark[1]}$ &
      $\stackrel{V_{\tiny{\mathrm{O-O,br-hol}}}}{0.08\footnotemark[1]}$ &
      \\

      $\stackrel{V_{\tiny{\mathrm{CO-CO,br-br}}}}{0.08\footnotemark[1]}$ &
      $\stackrel{V_{\tiny{\mathrm{CO-CO,hol-hol}}}}{0.13\footnotemark[1]}$ &
      $\stackrel{V_{\tiny{\mathrm{CO-CO,br-hol}}}}{0.14\footnotemark[1]}$ &
      \\

      $\stackrel{V_{\tiny{\mathrm{O-CO,br-br}}}}{0.06\footnotemark[1]}$ &
      $\stackrel{V_{\tiny{\mathrm{O-CO,hol-hol}}}}{0.11\footnotemark[1]}$ &
      $\stackrel{V_{\tiny{\mathrm{O-CO,br-hol}}}}{0.13\footnotemark[1]}$ &
      $\stackrel{V_{\tiny{\mathrm{O-CO,hol-br}}}}{0.12\footnotemark[1]}$ \\

      \multicolumn{4}{l}{PdO $\sqrt{5}$ diffusion barriers} \\
      \hline
      $\stackrel{\rm{CO,br}\rightarrow\rm{br}}{0.4\footnotemark[1]}$ &
      $\stackrel{\rm{CO,hol}\rightarrow\rm{hol}}{0.6\footnotemark[1]}$ &
      $\stackrel{\rm{CO,br}\rightarrow\rm{hol}}{0.3\footnotemark[1]}$ &
      \\

      $\stackrel{\rm{O,br}\rightarrow\rm{br}}{1.2\footnotemark[1]}$ &
      $\stackrel{\rm{O,hol}\rightarrow\rm{hol}}{1.4\footnotemark[1]}$ &
      $\stackrel{\rm{O,br}\rightarrow\rm{hol}}{0.1\footnotemark[1]}$ &
      \\

      \multicolumn{4}{l}{PdO$\sqrt{5}$ reaction barriers (LH)} \\
      \hline
      $\stackrel{E_{\rm{O br, CO br}}}{1.0\footnotemark[1]}$&
      $\stackrel{E_{\rm{O hol, CO hol}}}{1.6\footnotemark[1]}$&
      $\stackrel{E_{\rm{O br, CO hol}}}{0.5\footnotemark[1]}$&
      $\stackrel{E_{\rm{O hol, CO br}}}{0.9\footnotemark[1]}$\\

      \multicolumn{4}{l}{PdO$\sqrt{5}$ reaction barriers (ER)} \\
      \hline
      $\stackrel{E_{\rm{O hol 1}}}{0.8}$&
      $\stackrel{E_{\rm{O hol 2}}}{0.5}$&
      $\stackrel{E_{\rm{O br 1}}}{0.0}$&
      $\stackrel{E_{\rm{O br 2}}}{0.0}$\\

      \multicolumn{4}{l}{Pd(100) desorption barriers}\\
      \hline
      &
      $\stackrel{E^0_{\tiny{\mathrm{O,hollow}}}}{-1.25}$ &
      $\stackrel{E^0_{\tiny{\mathrm{CO,bridge}}}}{-1.93}$ &
      \\
      \multicolumn{4}{l}{Pd(100) diffusion barriers} \\
      \hline
      $\stackrel{\rm{CO,br}\rightarrow\rm{br}}{0.14}$ &
      $\stackrel{\rm{O,hol}\rightarrow\rm{hol}}{0.28}$ &
      \\
      \multicolumn{4}{l}{Pd(100) reaction barriers} \\
      \hline
      & & $\stackrel{E_{\rm{CO br, O hol}}}{0.9}$& \\
   \end{tabular}
   \end{ruledtabular}
   \footnotetext[1]{Rogal \emph{et al.}, PRB \textbf{77}, 155410-12, (2008)}
   \end{table}

Neglecting any lateral interactions beyond the short-range blocking rules, the required O and CO binding energies, diffusion and reaction barriers are independent of the local adsorbate environment. The corresponding values computed within our DFT setup are summarized in Table I and are generally in good agreement with previous computations \cite{zhang_accuracy_2007,rogal_first-principles_2007,todorova_pd1_2003}. The obtained LH reaction barrier of 0.9\,eV in particular is in good agreement with previous DFT calculations at varying coverages (0.76-1.05\,eV) \cite{hammer_no+co_2001, eichler_co_2002, zhang_co_2001-1}. Attempts to calculate an ER type reaction path over a $(2\times 2)$ oxygen adlayer at Pd(100) showed that CO and O coadsorption is energetically more favorable and thus an ER reaction is excluded from the Pd(100) model. As apparent from Table I the O and CO diffusion barriers are very low. This high mobility severely limits the numerical efficiency of the 1p-kMC simulations, which are completely dominated by frequent executions of diffusion events at minute time increments. In order to speed up the 1p-kMC simulations we thus artifically raised the diffusion barriers by as much as 0.5\,eV. As diffusion is then still fast enough to achieve full equilibration of the adlayer in the temperature range of interest to this study, we found this not to affect the computed coverages or TOFs at all.

\subsubsection{\texorpdfstring{$\sqrt{5}$}{sqrt5} surface oxide}

For the $\sqrt{5}$ surface oxide we employ the 1p-kMC model established and detailed by Rogal, Reuter, and Scheffler \cite{rogal_first-principles_2007,rogal_co_2008}. In brief, this model considers two non-equivalent sites named bridge and hollow as depicted on the right of Fig.~\ref{geom}. Since there is no process involving sites from adjacent bridge-hollow trenches, the lattice may be viewed as quasi one-dimensional. In analogy to the just described Pd(100) model, the elementary process list consists of all non-concerted adsorption, desorption, diffusion and LH reaction processes involving these sites. Dissociative O$_2$ adsorption requires two neighboring sites, and is only hindered by a sizable adsorption barrier of 1.9\,eV in the case of adsorption into two bridge sites. CO adsorption is unimolecular and not hindered by adsorption barriers.

As only modification of the Rogal model we additionally consider an ER reaction mechanism, as recently suggested by Hirvi \emph{et al.} for bulk PdO(101) \cite{hirvi_co_2010}. Reaction path calculations vertically impinging a CO molecule over O atoms adsorbed in the different surface sites indeed also yield rather low reaction barriers over the $\sqrt{5}$ surface oxide, namely about 0.7\,eV over hollow and essentially zero over bridge. The latter rather astonishing result has very little consequences for the surface oxide stability and the catalytic activity in near stoichiometric feeds though. Under corresponding gas-phase conditions the O coverage of bridge sites is negligible and ER reaction events involving O in bridge sites correspondinlgy made no significant contribution to the overall TOF in the simulations.

\subsection{kMC simulation setup}

Both 1p-kMC models were implemented using the kmos framework\cite{hoffmann_kmos_2012}. All simulations were performed in simulation cells containing $(20 \times 20)$ unit cells and using periodic boundary conditions. Systematic checks showed that the quantities of interest here, i.e. the average steady-state coverages and TOFs, are perfectly converged at these cell sizes. For defined gas-phase feed conditions $(T, p_{\mathrm{CO}}, p_{\mathrm{O}_2})$ the 1p-kMC simulations eventually reach a steady state, with constant TOF and average surface coverages $\bar{\Theta}_{i,st}$ of species $i$ on site $st$,
    \begin{equation}
    \bar{\Theta}_{i,st} = \frac{\sum_n \Theta_{i,st,n}\Delta t_{n}}{\sum_n\Delta t_{n}} \quad ,
    \end{equation}
where $n$ denotes the value at the $n$th kMC step after steady state has been reached. Starting from different initial adsorbate distributions (clean surface, O-poisoned surface and CO-poisoned surface) and using different random number seeds we validated that this steady state is well defined, i.e. we never observed multiple steady-states in this system. In the kinetic phase diagrams shown in Fig. \ref{fig_phasediag_600} below the obtained results are summarized in form of contour plots. All contour plots were interpolated using radial basis functions in order to reduce numerical noise.

\section{Results}

\subsection{Kinetic phase diagrams at 600\,K}

\begin{figure*}
\includegraphics[width=\textwidth]{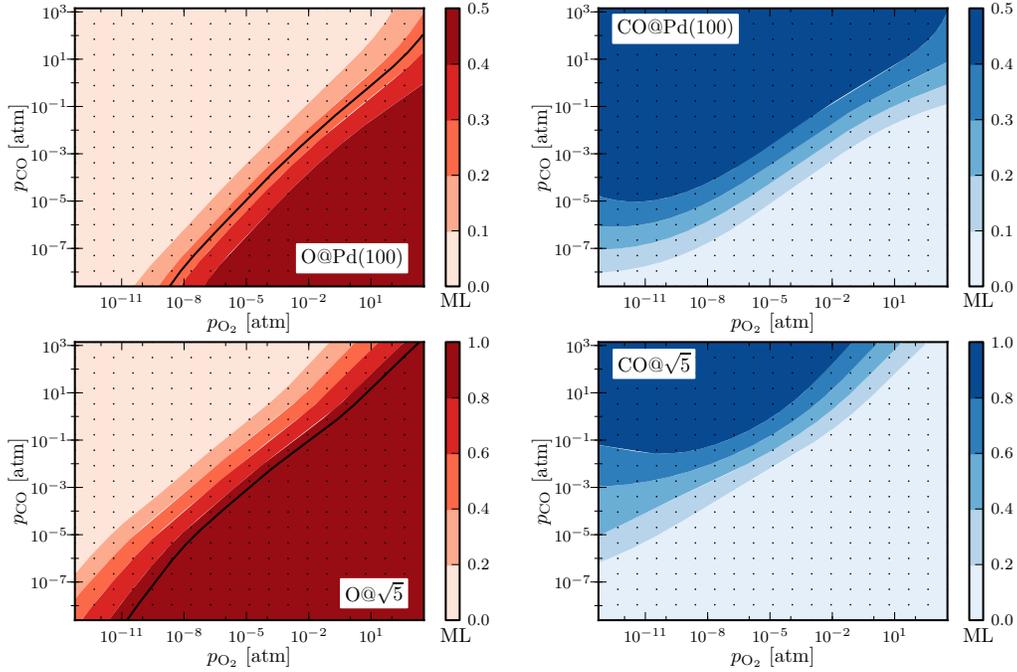}
\caption{\label{fig_phasediag_600}
(Color online) 1p-kMC computed O and CO coverage as a function of partial pressures for $T=600$~K at Pd(100) (upper panels) and at the $\sqrt{5}$ surface oxide (lower panels). At Pd(100) the coverage refers to O in hollow sites and CO in bridge sites, at the $\sqrt{5}$ the coverage refers to O and CO in upper hollow sites (see text). The thick black line in the left panels indicates the boundary where each phase is expected to be stable: For Pd(100) this corresponds to the $p\left(2\times2\right)$ oxygen adlayer, for PdO$\sqrt{5}$ this is a coverage $>0.9$~ML.}
\end{figure*}

Predicted oxygen and CO coverages on the $\sqrt{5}$, as well as on the Pd(100) surface are shown in Fig. \ref{fig_phasediag_600} for a temperature of 600\,K. For the Pd(100) surface the coverages refer to CO in bridge sites and O in hollow sites, as these are the only available sites for these species in the employed model, respectively. For the $\sqrt{5}$ surface only the O and CO coverage in the upper hollow sites is shown, as this is the expected critical descriptor for the stability of the surface oxide layer \cite{rogal_first-principles_2007}. For Pd(100) a coverage of 1\,monolayer (ML) corresponds to one adsorbate (O or CO) per surface Pd atom. For the $\sqrt{5}$ surface a coverage of 1\,ML corresponds to a complete occupation of the upper hollow rows by the respective species.

Both models yield the intuitively expected coverage variations with partial pressure. Starting with pure O$_2$ gas conditions (going horizontally along the bottom of the panels in Fig. \ref{fig_phasediag_600}), both models yield the correct surface coverages in the respective stable regimes by construction: For Pd(100) increasing O$_2$ content in the gas phase leads to a gradual coverage increase starting from the clean surface. For the $\sqrt{5}$ the full ML O coverage in the bottom right part of the panel in turn reflects a fully intact layer of the surface oxide in which all upper hollow sites are fully covered by oxygen. Increasing CO pressure also influences the surface coverages as intuitively expected: An increasing CO population at the surface defers the stabilization of surface O at Pd(100) to higher O$_2$ pressures, while it leads to a quicker depletion of upper hollow O atoms at the $\sqrt{5}$ at decreasing O$_2$ pressures.
For highly overstoichiometric CO pressures (upper horizontal line in the panels) the Pd(100) model displays CO coverages of 0.5~ML. This is less than the highest CO concentration characterized experimentally, namely 0.75~ML \cite{tushaus_understanding_1990}. This deviation results from the employed simplified lateral interaction model, which likely features too repulsive CO-CO interactions. We will scrutinize this with the less repulsive Pd-1NN model below, but also note that these CO-rich gas-phase conditions (and dense CO adlayers) are not the focus of our present interest. The same holds for the incorrect limit of a purely CO-covered $\sqrt{5}$ surface in the upper left part of the respective panel (where the surface oxide would in reality be reduced away), as well as for the $c(2 \times 2)$ 0.5\,ML O coverage in the lower right part of the Pd(100) panel (where instead a surface or bulk oxide would be formed).

\begin{figure}
\includegraphics[width=0.6\columnwidth]{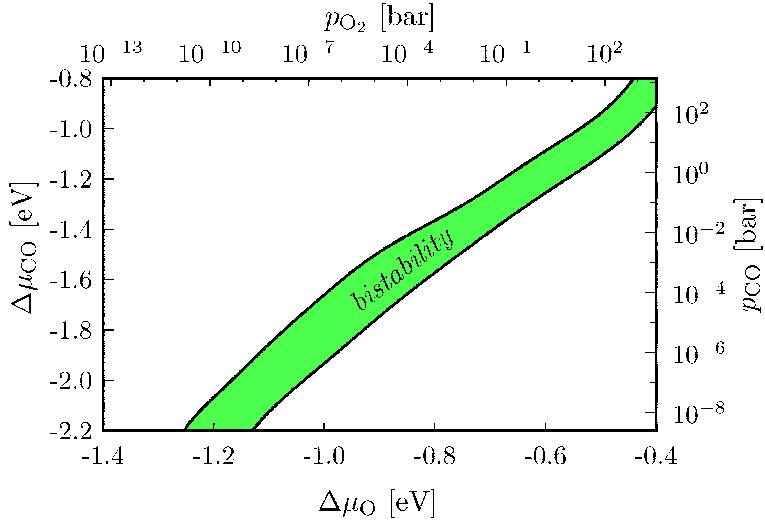}
\caption{\label{bistability_600K}
(Color online) Deduced region of bistability, i.e. gas-phase conditions where the 1p-kMC models would predict the simultaneous stability of pristine Pd(100) and the $\sqrt{5}$ surface oxide (see text). All diagram show
the same range of gas phase chemical potentials.}
\end{figure}

The very transition between CO- and O-covered regimes requires closer inspection. In the case of the Pd(100) surface a CO-poisoned surface would be catalytically inactive, but certainly stable. This surface is expected to be stable at least until the oxygen concentration does not rise above $\sim 0.25$~ML as this corresponds to a coverage regime representative for the experimentally characterized $p\left(2\times 2\right)$ overlayer \cite{lundgren_kinetic_2004,zheng_oxidation_2002}. In Fig. \ref{fig_phasediag_600} we therefore denote the stability boundary of the Pd(100) surface at 0.25\,ML coverage, i.e. we would expect a stable Pd(100) surface for any gas-phase conditions to the upper left of this line. Note that due to the steep coverage rise in the transition region, this stability boundary would be barely affected on the scale of Fig. \ref{fig_phasediag_600} if we had e.g. chosen 0.4\,ML coverage as the stability criterion. In the same spirit Rogal \emph{et al.} have used a coverage exceeding 0.9\,ML as stability criterion for the $\sqrt{5}$ surface oxide before \cite{rogal_co_2008}. The corresponding stability boundary for the surface oxide is also drawn in Fig. \ref{fig_phasediag_600}, and we would expect the surface oxide to be stable anywhere to the bottom right of this line. Again, on the scale of Fig. \ref{fig_phasediag_600} it would make little difference, if a stability criterion of e.g. 0.95\,ML or 0.99\,ML coverage had been used. With the two stability regions thus quite narrowly defined, the central and intriguing feature of Fig. \ref{fig_phasediag_600} is that there is a finite range of partial pressures where both models are predicted to be stable. This is highlighted again in Fig. \ref{bistability_600K}, where the resulting bistability region is marked in green.

\subsection{Variation with temperature}

The results presented in the preceding section were obtained for $T = 600$\,K, a temperature that falls in the middle of the temperature range from $\sim 300\rm{-} 800$\,K that is generally most relevant for CO oxidation catalysis. Not least to make contact with the dedicated reactor STM experiments performed by Hendriksen {\em et al.} at 408-443\,K \cite{hendriksen_bistability_2005}, an important next step is to assess the variation of our findings, in particular the existence of a bistability region, with temperature. Correspondingly and following the same protocol as before, the two lower panels of Fig. \ref{bistability_600K} summarize our findings for $T = 500$\,K and $T = 400$\,K. Both panels show the same range of gas-phase chemical potentials as the upper panel for 600\,K, allowing to ascribe all differences between the three panels directly to kinetic effects, that is deviations from thermodynamic scaling. Intriguingly, these deviations are rather small, i.e. the differences for the three temperatures amount primarily to the expected scaling of the pressure axes \cite{reuter_first-principles_2006}. This leaves the central result in form of a bistability over a finite range of gas-phase pressures essentially untouched within the temperature range 400-600\,K. Particularly intriguing is that certainly for $T = 600$\,K the catalytically most relevant near-ambient conditions at stoichiometric feed fall right within the region of bistability. For the lower temperatures the bistability region shifts increasingly towards O-rich conditions, such that e.g. the conditions of the Hendriksen experiments fall just at the border of this region.

\section{Discussion}

\subsection{Origin and robustness of bistability region}

\begin{figure}
\includegraphics[width=\columnwidth]{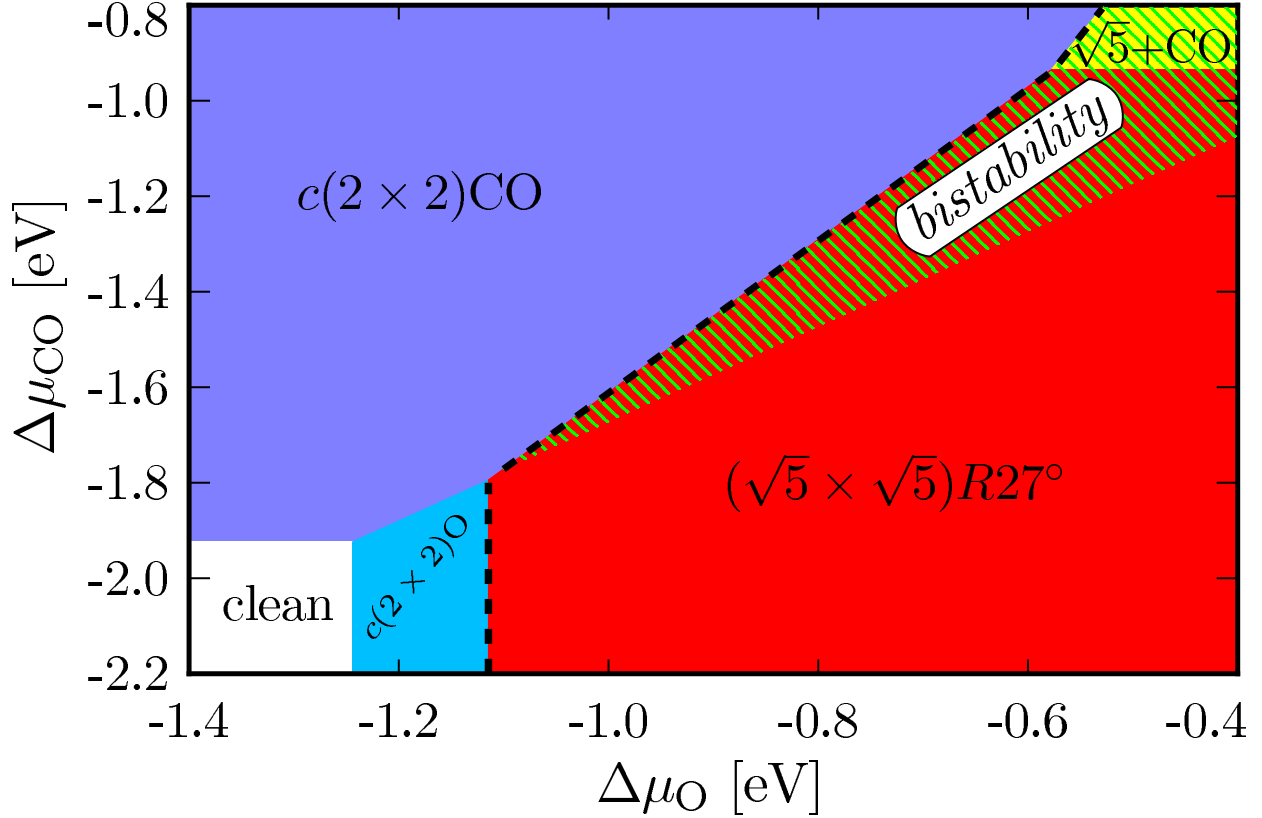}
\caption{\label{thermodynamic_bistability}
(Color online) "Constrained" {\em ab initio} thermodynamic phase diagram of the Pd(100)/$\sqrt{5}$ system, considering only ordered structures consistent with the 1p-kMC lateral interaction models (see text). The thick dashed line denotes the stability boundary between adsorption phases on Pd(100) (upper left part) and adsorption phases on the $\sqrt{5}$ surface oxide (bottom right part). Additionally shown as hatched area is the stability region of Pd(100) phases with O coverage below 0.5\,ML. This is the thermodynamic equivalent to the bistability region identified in the 1p-kMC simulations, if the low O coverage defers formation of the (thermodynamically preferred) surface oxide.}
\end{figure}

An immediate concern with the observed bistability region is that it is rather narrow in $(p_{\rm O_2},p_{\rm CO})$-space. Considering the range of uncertainties underlying the 1p-kMC models, prominently the semi-local DFT energetics and lateral interaction model, this raises doubts as to the robustness of this finding. Fortunately, in the present case an analysis of the atomic-scale reason behind the bistability shows that its actual existence emerges rather independently from these uncertainties, which is why we lead these two discussions jointly in this section.

A first important step towards an understanding of the atomic-scale origin of the bistability comes from the observed almost perfect thermodynamic scaling of the location and extent of the bistability region in $(p_{\rm O_2}, p_{\rm CO}$)-space in the temperature range 400-600\,K, cf. Fig. \ref{bistability_600K}. This suggests that the actual reaction kinetics, and the uncertainties in the concomitant reaction barriers, is not central to its existence. This view is confirmed by the fact that an equivalent bistability region is already obtained within a "constrained" {\em ab initio} thermodynamics approach \cite{reuter_first-principles_2003,reuter_2003}, i.e. an approach that neglects the reaction kinetics completely. Figure \ref{thermodynamic_bistability} shows the corresponding phase diagram for the same range of chemical potentials also underlying the panels in Fig. \ref{bistability_600K}. For a comparison to the 1p-kMC results, exactly the same DFT energetics and only ordered structures consistent with the 1p-kMC lateral interaction models are used. This explains small differences with respect to the corresponding phase diagram published before by Rogal, Reuter, and Scheffler \cite{rogal_first-principles_2007}, which e.g. considered additional experimentally characterized CO ordered phases (not of relevance for the present discussion). Denoted by the thick dashed line is the stability boundary between adsorption phases on the pristine Pd(100) surface (upper left part) and adsorption phases on the $\sqrt{5}$ surface oxide (lower right part). For all catalytically relevant gas-phase conditions, this boundary runs in fact between CO-covered Pd(100) and different adsorption phases on the $\sqrt{5}$, i.e. with increasing O-content in the gas phase constrained thermodynamics predicts an abrupt phase transition from a CO-poisoned surface directly to the surface oxide. Knowing that a critical local coverage of about $\sim 0.5$\,ML O is necessary to induce the formation of the surface oxide, it is therefore interesting to also include in the phase diagram the stability region of O-containing Pd(100) phases with less than 0.5\,ML O coverage, i.e. phases where (coming from the pristine metal side) oxide formation would not yet start. Intriguingly, the resulting region shown in Fig. \ref{thermodynamic_bistability} extends over similar gas-phase conditions as the bistability region deduced from the 1p-kMC simulations.

As such we ascribe the bistability region to gas-phase conditions, where the $\sqrt{5}$ surface oxide is thermodynamically more stable, but where a too low O coverage at Pd(100) would not readily induce the formation of the oxide. When crossing the bistability region from O-rich to CO-rich gas-phase conditions the system will thus prevail in the (thermodynamically preferred) oxidized state, while in the opposite direction it will prevail in the metal state as kinetic limitations to stabilize enough oxygen at the surface prevent the formation of the surface oxide. With this understanding of the atomic-scale origin, it is primarily the adsorbate binding energies and the lateral interaction model that are crucial to the robustness of the bistability region -- as they govern the stabilization of oxygen at the surface. More specifically, it is the relative binding energy differences at the metal and the $\sqrt{5}$ that will primarily affect the extension of the bistability region, while the absolute binding energetics will rather shift its location in $(p_{\rm O_2}, p_{\rm CO})$-space. We would expect uncertainties in the employed DFT exchange-correlation functional to rather affect the absolute binding energetics (i.e. systematic over- or underbinding) and only to a lesser extent the binding energy differences of O and CO at the two surfaces. This view is confirmed by systematic tests, in which we increased or decreased all binding energies by 0.2\,eV and in both cases found only small differences in the extension of the bistability region at more pronounced changes in its position in $(p_{\rm O_2}, p_{\rm CO})$-space. Correspondingly, we do not expect that the predicted gas-phase conditions for the bistability are accurate to better than some orders of magnitude in pressure. Its actual existence, however, should be very robust with respect to the uncertainties of present-day DFT functionals.

\begin{figure}
\includegraphics[width=\columnwidth]{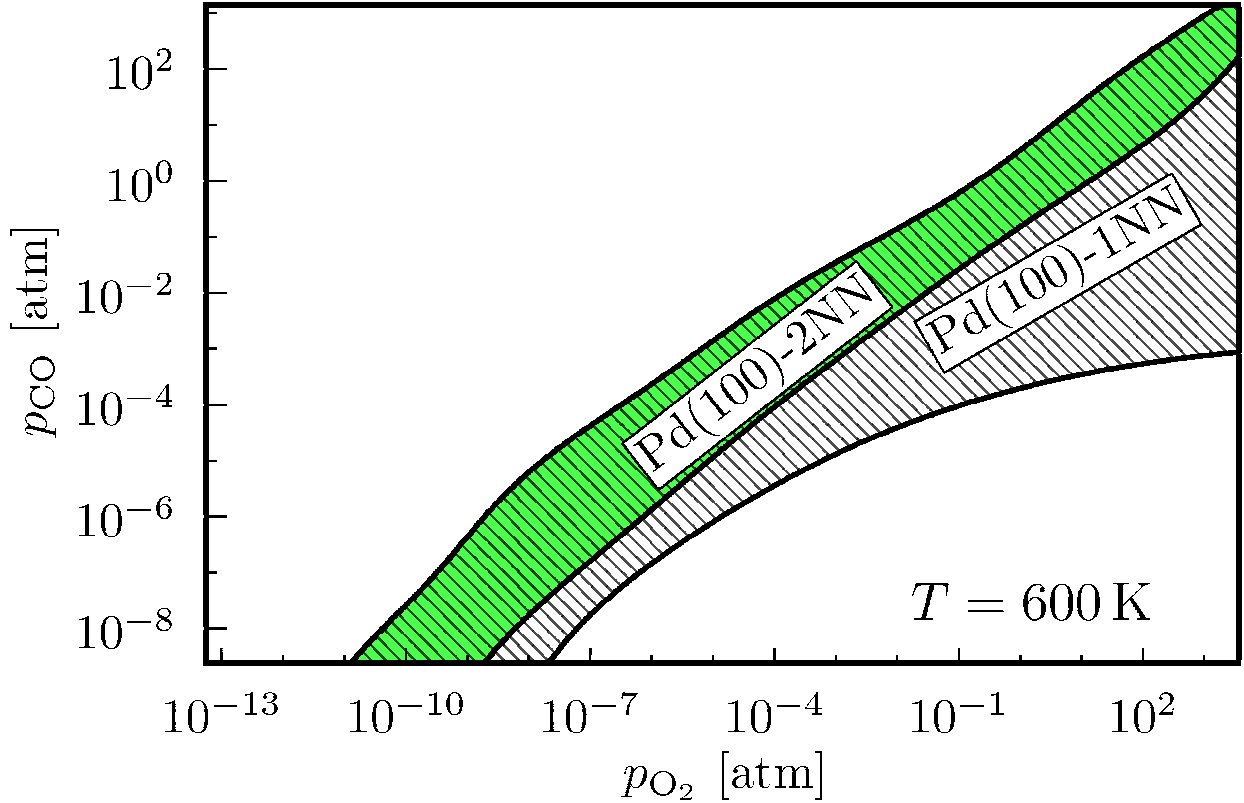}
\caption{\label{phase_boundaries}
(Color online) Comparison of the bistability regions arising from 1p-kMC simulations with different lateral interaction models. Pd(100)-2NN is the default model employed sofar, where configurations are excluded that would result in CO adsorbed in nearest-neighbor and next-nearest neighbor positions (green solid area). Pd(100)-1NN mimics less repulsive interactions and only excludes configurations resulting in CO adsorbed in nearest-neighbor positions (hatched area).}
\end{figure}

This leaves as final important aspect the employed lateral interaction model on the Pd(100) surface. Already the comparison to the experimentally determined maximum CO coverage (0.75\,ML \cite{tushaus_understanding_1990} versus 0.5\,ML in Fig. \ref{fig_phasediag_600}) indicated the site-blocking rules in the default Pd(100)-2NN model to be too repulsive. We assess the consequences for the bistability region by determining this region with the less repulsive Pd(100)-1NN model, where now only configurations that would result in CO adsorbed in nearest-neighbor positions are excluded. The resulting 1p-kMC bistability region at 600\,K is shown in Fig. \ref{phase_boundaries} and compared to the corresponding bistability region deduced before within the default Pd(100)-2NN model. Obviously the eased stabilization of CO at the Pd(100) surface (maximum coverage now 1\,ML) blocks the O adsorption even more effectively and consequently leads to a much increased range of gas-phase conditions where both surface states are predicted to be stable. If we take the comparison to the maximum CO coverage from experiment as measure, we would expect the true CO-CO repulsion to be somewhere between the one represented within the Pd(100)-1NN and Pd(100)-2NN site blocking models. Correspondingly, we would estimate the true extension of the bistability region to be somewhere between those predicted within the two models. While thus not fully quantified in its extension, the actual existence of a bistability region {\em per se} is robust against the uncertainty in the employed lateral interaction model.

\subsection{Comparison to experiment}

The finding of a finite bistability region in our calculations agrees nicely with the bistability and oscillations reported experimentally by Hendriksen and coworkers \cite{hendriksen_oscillatory_2004, hendriksen_bistability_2005}. Unfortunately, we can not directly compare the temperatures and pressures where this bistability is found. On the theoretical side this is due to the aforediscussed uncertainty with respect to the location of this bistability region in $(p_{\rm O_2},p_{\rm CO})$-space caused by the approximate first-principles parameters and lateral interactions entering the kinetic models. On the experimental side this is due to the suspected non-negligible mass-transfer effects in the employed reactor STM setup \cite{van_rijn_surface_2011,matera_2009,matera_2010}. Nevertheless, the deduced pressure ranges are intriguingly close, as is the extension of the bistability region. At the experimental $T\sim 400$~K and $p_{\rm O_2}=1$\,atm, the measured width of the bistability region in CO pressure is 0.02\,atm \cite{hendriksen_bistability_2005}, while it is $\approx0.5$\,atm in the simulations. Also, the finding that at constant $p_{\rm O_2}$, the bistability region shifts with increasing temperature to increasing $p_{\rm CO}$ agrees with the experimental observations \cite{hendriksen_oscillatory_2004}. As such our interpretation is that the here obtained bistability between metallic Pd(100) and the $\sqrt{5}$ surface oxide provides an atomic-scale model for the oscillations of the Hendriksen experiments. This is close to the interpretation arrived at in the experimental study, with the slight modification that they suspected the metallic phase to correspond to predominantly O-covered Pd(100), whereas our simulations show that this is predominantly CO-covered Pd(100).

\begin{figure}
\includegraphics[width=0.8\columnwidth]{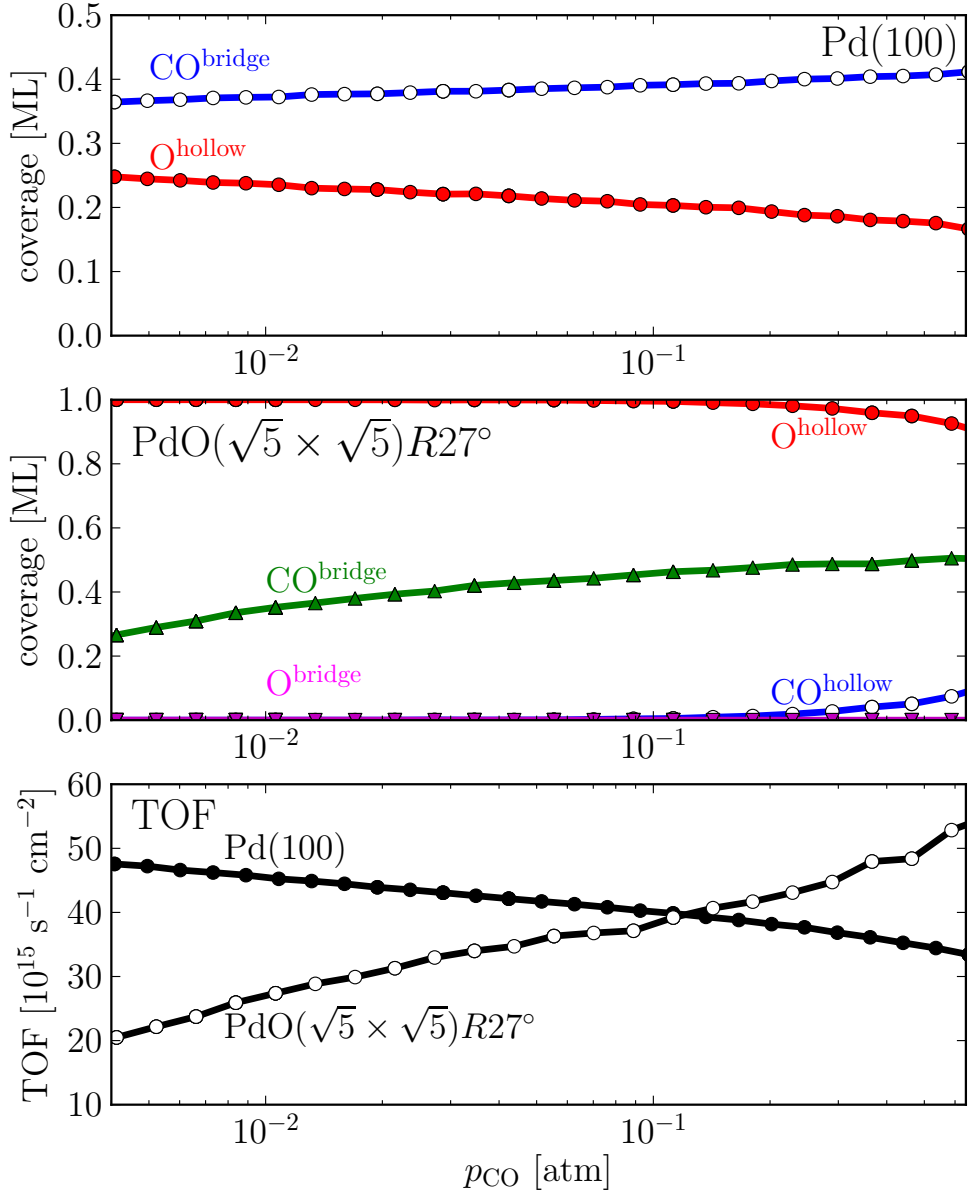}
\caption{\label{formal_kinetics}
(Color online) 1p-kMC surface coverages and turn-over-frequencies (TOFs) at gas-phase conditions comparable to the reactor STM experiments by Hendriksen {\em et al.} \cite{hendriksen_bistability_2005}. Shown is the bistability region in $p_{\rm CO}$ at fixed $T = 400$\,K and $p_{\rm O_2} = 1$\,atm. Upper panel: O and CO coverage on the Pd(100) surface; middle panel: O and CO coverage on the $\sqrt{5}$ surface oxide; lower panel: comparison of the intrinsic TOFs of the two system states.}
\end{figure}

The atomic-scale insight into the surface composition provided by our 1p-kMC simulations also allows to further qualify the reaction mechanism. An intriguing observation made by Hendriksen and coworkers was that in contrast to the metal surface, the reaction rate on the oxide did not follow traditional LH kinetics. They proposed this as a signature of low oxide stability and a concomitant Mars-van-Krevelen type mechanism, in which the oxide is continually consumed and reformed. Figure \ref{formal_kinetics} shows that this can instead be rationalized within a much simpler picture that does not involve the oxide stability. Shown are the surface coverages on the Pd(100) metal and $\sqrt{5}$ surface oxide over the $p_{\rm CO}$ pressure range, where we obtain bistability at fixed $T = 400$\,K and $p_{\rm O_2} = 1$\,atm, i.e. for gas-phase conditions comparable to the Hendriksen experiments. At the Pd(100) surface, O and CO adsorb in inequivalent sites, hollow and bridge, respectively. Due to the strongly repulsive interactions, mimicked in our simulations by mutual site blocking, this nevertheless leads effectively to a competition for adsorption sites between the two adsorbates and in consequence to a LH-type kinetics. In contrast, on the $\sqrt{5}$ surface oxide, O and CO also adsorb in inequivalent sites at these gas-phase conditions, namely O almost exclusively on the hollow sites and CO almost exclusively on the bridge sites, cf. Fig. \ref{formal_kinetics}. At the surface oxide these sites are located further away from each other and adsorbates even in nearest-neighbor sites experience only very small lateral interactions, cf. Table I. Consequently, the occupation of these two site types occurs almost independently of each other at varying gas-phase conditions and gives rise to a TOF proportional in CO pressure, cf. Fig. \ref{formal_kinetics}, exactly as observed in the experiment. Similarly, we find the TOF to vary only little when changing the oxygen partial pressure away from the conditions shown in Fig. \ref{formal_kinetics} -- again fully consistent with the experimental findings.

The understanding that the actual reaction mechanism over the $\sqrt{5}$ proceeds by adsorption on the otherwise intact surface oxide suggests that the continued roughening of the surface observed in the experiments is a by-product and not essential to the catalytic activity. With respect to the total activity Fig. \ref{formal_kinetics} shows that over the bistability region both surface states, Pd(100) and $\sqrt{5}$, exhibit rather similar intrinsic TOFs to within a factor of three. While the mass-transfer limitations present in the reactor STM measurements prevent a direct comparison, this finding alone sets the controversial discussion concerning {\em the} active state of the surface into perspective. At low temperatures and UHV conditions, our 1p-kMC models indeed yield a significantly higher activity of the Pd(100) surface. This arises predominantly from the comparatively weak CO binding at the $\sqrt{5}$ surface oxide and the concomitant limitations in stabilizing it at the surface. However, at technological gas-phase conditions, which in our simulations do fall within the bistability region, the TOF differences between the two models are not large enough and furthermore change sensitively with the detailed gas-phase conditions, cf. Fig. \ref{formal_kinetics}, to support a detailed discussion as to which state is the more active one. In this respect, the Pd(100) surface differs qualitatively from the equally prominently discussed Ru/RuO$_2$ system \cite{reuter_review,over_2003}. On ruthenium the catalytic activity can be clearly attributed to the oxidized surface, while in the present system two competing surface states can equally contribute to the catalytic activity, which thus also explains why a clear signature of palladium (surface) oxide formation may in some circumstances be more difficult to find.

\section{Conclusions}

Detailed 1p-kMC simulations on either the Pd(100) surface or the $\sqrt{5}$ surface oxide yield a finite range of CO oxidation gas-phase conditions, where both surface states appear stable. This finding of a bistability region is robust against variations in the detailed criteria used to assess the stability of either state, as well as against the uncertainties arising from the approximate first-principles energetics and lateral interaction models. Notwithstanding the latter uncertainties do affect the position where in $(T,p)$-space the bistability region is found and we expect this to translate in uncertainties up to the order of 100\,K and some orders of magnitude in pressure.

The bistability region arises from limitations in stabilizing oxygen at the Pd(100) surface, which extends the stability region of the metallic surface beyond that predicted by thermodynamics. In some respect this is therefore reminiscent of the generic Langmuir-Hinshelwood kinetics that lead to bistability between a CO-poisoned state that effectively blocks O$_{2}$ adsorption and an O-rich state, i.e. the bistability of the reaction rate originates essentially from the inequivalence of the adsorption of carbon monoxide and oxygen.\cite{sales_oscillatory_1982,bykov_steady_1981,bar_theoretical_1992,zhdanov_kinetic_1994,eiswirth_oscillating_1996}. The difference is that in these classical models the two stable solutions were either supposed to be different adsorbate phases on the same substrate or did correspond to an active metal and an inactive oxide state. Instead, in the present system we have a CO-rich Pd(100) state that blocks oxygen adsorption and the surface oxide as O-rich state. Furthermore, the bistability does comprise technological conditions and conditions comparable to the dedicated reactor STM experiments performed by Hendriksen {\em et al.}\cite{hendriksen_bistability_2005,hendriksen_role_2010}. Under these conditions, both surface terminations do show similar intrinsic activity, which sets preceding discussions with respect to {\em the} active state into perspective.

Within the uncertainties of the exact location of the bistability region in $(T,p)$-space we suggest the here obtained bistability between metallic Pd(100) and the $\sqrt{5}$ surface oxide as an atomic-scale model for the oscillations of the Hendriksen experiments. Likewise, our data lends further support to the view that oxide formation plays an important role in understanding the catalytic activity of Pd catalysts. If it is even the oscillations themselves, catalytic activity would again be rationalized as a kinetic phase transition phenomenon, i.e. the catalyst surface being close to an instability\cite{ziff_kinetic_1986,evans_kinetic_1991,reuter_kMC}. At present our simulations performed separately on the two intact surface states, corresponding to ideal terraces of Pd(100) and the $\sqrt{5}$ surface oxide, can not address this notion directly. This holds equally for the experimental interpretation that the continued roughening of the oxidic surface during reaction, as well as the formation of steps are crucial ingredients to the oscillatory behavior. Our ongoing work is directed to extend the 1p-kMC capabilities towards the actual oxide formation and decomposition, which will then allow us to start scrutinizing these points with predictive-quality theory.

\section{Acknowledgements}
We gratefully acknowledge support from the German Research Council (DFG) and the TUM Faculty Graduate Center Chemistry, as well as generous computing time at the Supercomputing Center of the Max-Planck-Society, Garching. We thank Matthias Scheffler for insightful discussions and his continued support for this project.

\bibliographystyle{prsty}

\begin{thebibliography}{10}

\bibitem{gandhi_automotive_2003}
H.S. Gandhi, G.W. Graham, and R.W. {McCabe}, J. Catal. {\bf 216}, 433  (2003).

\bibitem{gao_reply_2010}
F. Gao, Y. Wang, and D.W. Goodman, J. Phys. Chem. C {\bf 114},  6874  (2010).

\bibitem{van_rijn_comment_2010}
R. van Rijn, O. Balmes, R. Felici, J. Gustafson, D. Wermeille, R. Westerstr\"om, E. Lundgren and J.W.M. Frenken, J. Phys. Chem. C {\bf 114}, 6875 (2010).

\bibitem{hirvi_co_2010}
J.T. Hirvi, T.-J.J. Kinnunen, M. Suvanto, T.A. Pakkanen, and J.K. N{\o}rskov, J. Chem. Phys. {\bf 133}, 084704 (2010).

\bibitem{zheng_reactivity_2002}
G. Zheng and E.I. Altman, J. Phys. Chem. B {\bf 106}, 1048 (2002).

\bibitem{chang_oxygen_1988}
S.-L. Chang and P.A. Thiel, J. Chem. Phys. {\bf 88}, 2071 (1988).

\bibitem{todorova_pd1_2003}
M. Todorova, E. Lundgren, V. Blum, A. Mikkelsen, S. Gray, M. Borg, J. Gustafson, J. Rogal, K. Reuter, J.N. Andersen, and M. Scheffler, Surf. Sci. {\bf 541}, 101  (2003).

\bibitem{kiss_deactivation_1985}
J.T. Kiss and R.D. Gonzalez, Ind. \& Eng. Chem. Prod. Res. and Development {\bf 24}, 216 (1985).

\bibitem{gabasch_comparison_2007}
H. Gabasch, A. Knop-Gericke, R. Schl\"ogl, M. Borasio, C. Weilach, G. Rupprechter, S. Penner, B. Jenewein, K. Hayek, and B. Kloetzer,
Phys. Chem. Chem. Phys. {\bf 9}, 533 (2007).

\bibitem{gao_infrared_2008}
F. Gao, M. Lundwall, and D.W. Goodman, J. Phys. Chem. C {\bf 112}, 6057 (2008).

\bibitem{lundgren_kinetic_2004}
E. Lundgren, J. Gustafson, A. Mikkelsen, J.N. Andersen, A. Stierle, H. Dosch, M. Todorova, J. Rogal, K. Reuter, and M. Scheffler,
Phys. Rev. Lett. {\bf 92}, 046101 (2004).

\bibitem{rogal_first-principles_2007}
J. Rogal, K. Reuter, and M. Scheffler, Phys. Rev. Lett. {\bf 98}, 046101  (2007).

\bibitem{rogal_co_2008}
J. Rogal, K. Reuter, and M. Scheffler, Phys. Rev. B {\bf 77}, 155410 (2008).

\bibitem{hendriksen_oscillatory_2004}
B. Hendriksen, S. Bobaru, and J. Frenken, Surf. Sci. {\bf 552}, 229 (2004).

\bibitem{hendriksen_bistability_2005}
B. Hendriksen, S. Bobaru, and J. Frenken, Catal. Today {\bf 105}, 234 (2005).

\bibitem{hendriksen_role_2010}
B.L.M. Hendriksen, M.D. Ackermann, R. van Rijn, D. Stoltz, I. Popa, O. Balmes, A. Resta, D. Wermeille, R. Felici, S. Ferrer, and
J.W.M. Frenken, Nat. Chem. {\bf 2}, 730 (2010).

\bibitem{rogal_co_2007}
J. Rogal, K. Reuter, and M. Scheffler, Phys. Rev. B {\bf 75}, 205433 (2007).

\bibitem{zhang_accuracy_2007}
Y. Zhang, V. Blum, and K. Reuter, Phys. Rev. B {\bf 75}, 235406 (2007).

\bibitem{behm_adsorption_1980}
R.J. Behm, K. Christmann, G. Ertl, and M.A. Van~Hove, J. Chem. Phys. {\bf 73}, 2984 (1980).

\bibitem{liu_chemical_2006}
D.-J. Liu and J.W. Evans, J. Chem. Phys. {\bf 125}, 054709 (2006).

\bibitem{liu_atomistic_2009}
D.-J. Liu and J.W. Evans, Surf. Sci. {\bf 603}, 1706 (2009).

\bibitem{temel_2007}
B. Temel, H. Meskine, K. Reuter, M. Scheffler, and H. Metiu, J. Chem. Phys. {\bf 126}, 204711 (2007).

\bibitem{reuter_2011_book}
K. Reuter, in {\em Modeling and Simulation of Heterogeneous Catalytic Reactions: From the Molecular Process to the Technical System}, Ed. O. Deutschmann, Wiley-VCH, Weinheim (2011), pp. 71-112.

\bibitem{sabbe_2012}
M.K. Sabbe, M.-F. Reyniers, and K. Reuter, Catal. Sci. Technol. {\bf 2}, 2010 (2012).

\bibitem{reuter_first-principles_2006}
K. Reuter and M. Scheffler, Phys. Rev. B {\bf 73}, 045433 (2006).

\bibitem{reuter_composition_2001}
K. Reuter and M. Scheffler, Phys. Rev. B {\bf 65}, 035406 (2001).

\bibitem{chase_jr_nist-janaf_1998}
M.W. Chase~Jr, J. Phys. Chem. Ref. Data {\bf 9}, 1 (1998).

\bibitem{henkelman}
H. Jonsson, G. Mills, and K.W. Jacobson, {\em Nudged Elastic Band Method for Finding Minimum Energy Paths of Transitions, in Classical and Quantum Dynamics in Condensed Phase Simulations}, B.J. Berne, G. Cicotti, and D.F. Coker (Eds.), World Scientific, New Jersey (1998).

\bibitem{henkelman2}
G. Henkelman, B.P. Uberuaga, and H. Jonsson, J. Chem. Phys. {\bf 113}, 9901 (2000).

\bibitem{kostelnik}
P. Kostelnik, N. Seriani, G. Kresse, A. Mikkelsen, E. Lundgren, V. Blum, T. Sikola, P. Varga, and M. Schmid, Surf. Sci. {\bf 601}, 1574 (2007).

\bibitem{perdew_generalized_1996}
J.P. Perdew, K. Burke, and M. Ernzerhof, Phys. Rev. Lett. {\bf 77}, 3865 (1996).

\bibitem{clark_first_2005}
S.J. Clark, M.D. Segall, C.J. Pickard, P.J. Hasnip, M.I.J. Probert, K. Refson, and M.C. Payne, Z. Kristallogr. {\bf 220} 567 (2009)

\bibitem{stuve_co_1984}
E. Stuve, R. Madix, and C. Brundle, Surf. Sci. {\bf 146}, 155 (1984).

\bibitem{uvdal_structure_1988}
P. Uvdal, P.-A. Karlsson, C. Nyberg, S. Andersson, N.V. Richardson, Surf. Sci. {\bf 202}, 167 (1988).

\bibitem{tushaus_understanding_1990}
M. T{\"u}shaus, W. Berndt, H. Conrad, A.M. Bradshaw, and B. Persson, Appl. Phys. A {\bf 51}, 91 (1990).

\bibitem{rieder_helium_1985}
K. Rieder and W. Stocker, Surf. Sci. {\bf 150}, L66 (1985).

\bibitem{bradshaw_chemisorption_1978}
A.M. Bradshaw and F.M. Hoffmann, Surf. Sci. {\bf 72}, 513 (1978).

\bibitem{brundle_summary_1984}
C.R. Brundle, J. Behm, and J.A. Barker, J. Vac. Sci. Technol. A {\bf 2}, 1038 (1984).

\bibitem{chang_formation_1987}
S.L. Chang and P.~A. Thiel, Phys. Rev. Lett. {\bf 59}, 296 (1987).

\bibitem{liu_atomistic_2006}
D.-J. Liu and J.W. Evans, J. Chem. Phys. {\bf 124}, 154705 (2006).

\bibitem{meyer_first-principles_2011}
J. Meyer and K. Reuter, ({\em private communication}).

\bibitem{hammer_no+co_2001}
B. Hammer, J. Catal. {\bf 199}, 171 (2001).

\bibitem{eichler_co_2002}
A. Eichler, Surf. Sci. {\bf 498}, 314 (2002).

\bibitem{zhang_co_2001-1}
C.J. Zhang and P. Hu, J. Am. Chem. Soc. {\bf 123}, 1166  (2001).

\bibitem{hoffmann_kmos_2012}
M.J. Hoffmann, kmos, 2012, {\url{http://mhoffman.github.com/kmos}}.

\bibitem{zheng_oxidation_2002}
G. Zheng and E.I. Altman, Surf. Sci. {\bf 504}, 253 (2002).

\bibitem{reuter_first-principles_2003}
K. Reuter and M. Scheffler, Phys. Rev. Lett. {\bf 90}, 046103 (2003).

\bibitem{reuter_2003}
K. Reuter and M. Scheffler, Phys. Rev. B {\bf 68}, 045407 (2003).

\bibitem{van_rijn_surface_2011}
R. van Rijn, O. Balmes, A. Resta, D. Wermeille, R. Westerstr\"om, J. Gustafson, R. Felici, E. Lundgren, and J.W.M. Frenken,
Phys. Chem. Chem. Phys. 13, 13167 (2011).

\bibitem{matera_2009}
S. Matera and K. Reuter, Catal. Lett. {\bf 133}, 156 (2009).

\bibitem{matera_2010}
S. Matera and K. Reuter, Phys. Rev. B {\bf 82}, 085446 (2010).

\bibitem{reuter_review}
K. Reuter, Oil \& Gas Sci. and Technol. - Rev. IFP {\bf 61}, 471 (2006).

\bibitem{over_2003}
H. Over and M. Muhler, Prog. Surf. Sci. {\bf 72}, 3 (2003).

\bibitem{sales_oscillatory_1982}
B. Sales, J. Turner, and M. Maple, Surf. Sci. {\bf 114}, 381 (1982).

\bibitem{bykov_steady_1981}
V.I. Bykov, G.S. Yablonskii, and V.I. Elokhin, Surf. Sci. Lett. {\bf 107}, L334 (1981).

\bibitem{bar_theoretical_1992}
M. B{\"a}r, Ch. Z{\"u}licke, M. Eiswirth, and G. Ertl, J. Chem. Phys. {\bf 96}, 8595 (1992).

\bibitem{zhdanov_kinetic_1994}
V.P. Zhdanov and B. Kasemo, Surf. Sci. Rep. {\bf 20}, 113 (1994).

\bibitem{eiswirth_oscillating_1996}
M. Eiswirth, J. B{\"u}rger, P. Strasser, and G. Ertl, J. Phys. Chemistry {\bf 100},  19118  (1996).

\bibitem{ziff_kinetic_1986}
R.M. Ziff, E. Gulari, and Y. Barshad, Phys. Rev. Lett. {\bf 56}, 2553 (1986).

\bibitem{evans_kinetic_1991}
J.W. Evans, Langmuir {\bf 7}, 2514 (1991).

\bibitem{reuter_kMC}
K. Reuter, D. Frenkel, and M. Scheffler, Phys. Rev. Lett. {\bf 93}, 116105 (2004).

\end{thebibliography}

  \end{document}